\documentclass[12pt,preprint]{aastex}
\begin{document}

\title{Near Infrared Surface Properties of the Two Intrinsically
  Brightest Minor Planets (90377) Sedna and (90482) Orcus\footnote{Based on observations obtained at the Gemini
Observatory, which is operated by the Association of Universities for
Research in Astronomy, Inc., under a cooperative agreement with the
NSF on behalf of the Gemini partnership: the National Science
Foundation (United States), the Particle Physics and Astronomy
Research Council (United Kingdom), the National Research Council
(Canada), CONICYT (Chile), the Australian Research Council
(Australia), CNPq (Brazil), and CONICET (Argentina).}}

\author{Chadwick A. Trujillo}
\affil{Gemini Observatory, Northern Operations Center, 670 N A'ohoku
  Place, Hilo, Hawaii 96720}
\email{trujillo@gemini.edu}
\and
\author{Michael E. Brown}
\affil{California Institute of Technology, Division of Geological and
Planetary Sciences, MS 150-21, Pasadena, California 91125}
\email{mbrown@caltech.edu}
\and
\author{David L. Rabinowitz}
\affil{Yale Center for Astronomy and Astrophysics, Physics Department,
  Yale University, P.O. Box 208121, New Haven, Connecticut 06520-8121}
\email{david.rabinowitz@yale.edu}
\and
\author{Thomas R. Geballe}
\affil{Gemini Observatory, Northern Operations Center, 670 N A'ohoku
  Place, Hilo, Hawaii 96720}
\email{tgeballe@gemini.edu}

\author{\large Submitted to the Astrophysical Journal August 13, 2004}

\begin{abstract}
We present low resolution $K$ band spectra taken at the Gemini 8 meter
telescope of (90377) Sedna and (90482) Orcus (provisional designations
2003 $\rm VB_{12}$ and 2004 DW, respectively), currently the two minor
planets with the greatest absolute magnitudes (i.e. the two most
reflective minor planets).  We place crude limits on the surface
composition of these two bodies using a Hapke model for a wide variety
of assumed albedos.  The unusual minor planet (90377) Sedna was
discovered on November 14, 2003 UT at roughly 90 AU with 1.6 times the
heliocentric distance and perihelion distance of any other bound minor
planet.  It is the first solar system object discovered between the
Kuiper Belt and the Oort Cloud, and may represent a transition
population between the two.  The reflectance spectrum of (90377) Sedna
appears largely featureless at the current signal-to-noise ratio,
suggesting a surface likely to be highly processed by cosmic rays.
For large grain models (100 \micron\/ to 1 cm) we find that (90377)
Sedna must have less than 70\% surface fraction of water ice and less
than 60\% surface fraction of methane ice to $3 \sigma$ confidence.
Minor planet (90482) Orcus shows strong water ice absorption
corresponding to less than 50\% surface fraction for grain models 25
\micron\/ and larger.  Orcus cannot have more than 30\% of its surface
covered by large (100 mm to 1 cm) methane grains to $3 \sigma$
confidence.
\end{abstract}


\section{Introduction}

On November 14, 2003 UT, the minor planet (90377) Sedna (provisional
designation 2003 $\rm VB_{12}$) was found as part of an ongoing survey
for distant minor planets
\citep{2003EM&P...92...99T,2004ApJ...617..645B}.  As of this writing,
February 2005, with heliocentric distance 89.6 AU, (90377) Sedna is
the most distant body bound to the Sun by about a factor of 1.6
(cf. comet 35P/Herschel-Rigollet at 55.5 AU and Kuiper Belt object
2000 $\rm CR_{105}$ at 54.5 AU).  It also has the largest perihelion
distance (76 AU) of any solar system object by the same factor
(cf. 1999 $\rm CL_{119}$ with 47.8 AU perihelion and 2000 $\rm
CR_{105}$ with 44.3 AU).  The orbit of (90377) Sedna takes it from
perihelion (near its current location) out to about 1000 AU.
Formation scenarios for such an object are problematic; however, it is
most likely to have originated in our solar system and represents an
intermediate population between the Kuiper Belt and the Oort Cloud
\citep[and references therein]{2004ApJ...617..645B}.  The orbit of
(90482) Orcus (provisional designation 2004 DW) is more pedestrian as
it is in 3:2 resonance with Neptune, like Pluto.  Sedna and (90482)
Orcus currently are the two minor planets with the brightest absolute
magnitude ($H = 1.6$ and 2.4, respectively) and they have both been
discovered in the past year and a half.  Note that because of their
extreme distances, although these are likely to be among the largest
known minor planets, they are not the brightest by any means (Sedna is
$V \sim 21$ and (90482) Orcus is $V \sim 19$), requiring the use of
large telescopes for near-infrared investigations.  Here we present
near infrared spectra of both of their surfaces and place limits on
the presence of two volatiles: water ice and methane ice.

Due to its extreme distance, the surface of (90377) Sedna may be
considerably more pristine than any Kuiper Belt object (KBO), or any
closer object, which would be more susceptible to solar heating and
collisions.  The galactic cosmic ray environment of (90377) Sedna is
likely to be similar to that found in the Kuiper Belt.  Although large
objects such as (90377) Sedna may have considerable radiogenic thermal
processing deep in their interiors \citep{2002acm..conf...29M}, their
exteriors are likely to be minimally processed by solar heating, with
cosmic ray processing being the dominant source of surface
modification \citep{1987A&A...187..889J}.  The temperature range due
to solar heating of (90377) Sedna is 11 K -- 38 K between aphelion and
perihelion, lower than any known KBO (typically $\sim 45$ K).  In the
Kuiper Belt, collisions are thought to be rare between large bodies
\citep{1995AJ....110..856S}, but possibly significant in terms of
surface features given the extreme color dispersion seen for KBO
surfaces \citep{1996AJ....112.2310L}.  The collisional environment of
(90377) Sedna is much different than the typical KBO.  At aphelion,
where an object on an eccentric orbit spends the most time, the space
density of impactors and the likely velocity dispersion due to
Keplerian motion are very roughly $10^{4}$ and $10^{2}$ times lower
than in the Kuiper Belt, respectively.  Although numerical simulations
have produced collisions in the Oort Cloud
\citep{1988Icar...73..499S}, these simulations involve a large number
of very small ($\sim 1$ km) bodies, a population that has not been
observed in the Kuiper Belt and may be reduced from expected values
\citep{2003IAUJD..19E..47K,2003DPS....35.4903B}.  Thus, (90377) Sedna
has been minimally processed by solar heating and probably collisions
compared to the typical KBO.

The best constraints on the products of primitive volatiles bombarded
by high energy particles come from laboratory work, which suggests
that simple volatiles expected from ``pristine'' objects
(i.e. compositions unaltered since solar system formation) should
produce more complex organic materials after undergoing bombardment by
high energy photons and particles \citep{1983Icar...54..388M,
1987A&A...187..889J}.  Such bombardment causes the formation of
complex organics such as Titan tholins which are red throughout the
visible to the $J$-band (roughly 0.7 \micron\/ to 1.3 \micron\/).
Additional organics may also include primitive bitumens which are rich
in hydrocarbons and present transitions in the $K$-band due to
aromatic carbon-hydrogen (C-H) stretching, C=C stretching and
combinations of $\rm CH_{2}$ and $\rm CH_{3}$ stretching and symmetric
bending
\citep{1998Icar..134..253M,2002LPI....33.1525R,1994Icar..108..137M,2004Icar..168..158R}.
With additional bombardment, such compounds become more neutral and
dark as they lose hydrogen and become chemically closer to carbon
black \citep{1987A&A...187..889J}.  Additional compounds may also be
present, and become major components of primitive objects as seen on
Pholus which shows methanol, water, olivine, and tholins
\citep{1998Icar..135..389C}, Triton with methane, nitrogen, carbon
monoxide and carbon dioxide
\citep{1979ApJ...233.1016C,1984Icar...58..293C,1993Sci...261..742C,1999Icar..139..159Q},
Pluto with nitrogen, methane, and carbon monoxide
\citep{1976Sci...194..835C,1993Sci...261..745O,2000PASJ...52..551N},
and Kuiper Belt objects with water and possible metal-OH compounds
\citep{1999ApJ...519L.101B,2001AJ....122.2099J}.

Water ice has been observed on several other KBOs and is a major
constituent in comets.  Methane ice is considerably more rare in the
outer solar system, although it is thought that the largest, least
thermally evolved objects such as (90377) Sedna may retain it.
Recently, \cite{2004A&A...422L..43F} have reported a detection of
water ice on the surface of (90482) Orcus based on observations at the
3.56 m Telescopio Nazionale Galileo in La Palma.  Their results for
surface fraction covered by water ice are drawn from relatively lower
signal to noise near infrared spectra as well as optical spectra.
They have explored two possible models to explain their observations,
both of which assume quite low albedos for the surface (4\% and 10\%).
In this work, we place constraints on the presence of methane ice and
water ice on (90377) Sedna and (90482) Orcus using a simple Hapke
model computed for a wide variety of possible surface albedos and
grain diameters.

\section{Observations}

Reflectance spectra of (90377) Sedna and (90482) Orcus were collected
on UT 2003 December 27 and 2004 April 2 respectively from the
Frederick C. Gillett Gemini Observatory Northern 8-meter Telescope
using the Near InfraRed Imager and Spectrograph
\citep[NIRI,][]{2003PASP..115.1388H} in f/6 mode
(0.1165\arcsec/pixel).  Seeing for the two nights was stable
throughout the observations at approximately 0\farcs 6 to 0\farcs 7 in
$R$-band and 0\farcs 3 to 0\farcs 4 in $K$-band.  The Caltech
Submillimeter Observatory (CSO) reported a 225 GHz optical depth of
4\% to 7\% (corresponding to the driest quartile of Mauna Kea
conditions, approximately 1 to 1.7 mm precipitable water vapor)
throughout the observations of (90377) Sedna and 12\% to 16\%
(corresponding to approximately 3 to 4 mm precipitable water vapor)
for the observations of (90482) Orcus.  In total, 110 minutes of
on-source spectra were collected in $K$ band over an airmass of 1.03
to 1.21 for (90377) Sedna and 55 minutes were collected over an
airmass of 1.11 to 1.20 for (90482) Orcus.  Non-sidereal tracking
rates of roughly 1\arcsec/hr and 2\arcsec/hr were used to track
(90377) Sedna and (90482) Orcus, respectively.  The 0\farcs 7
spectroscopic slit was imaged before, during and after observation
sequences to verify that the objects were properly centered in the
slits.  For (90377) Sedna, solar analogue spectra of two G2V stars
were collected before the science sequence (1.13 airmasses, HD 377)
and after the science sequence (1.02 airmasses, HD 42807), with less
than 3\% variation between the two observed spectra throughout the
$K$-band, except in regions affected by telluric water and carbon
dioxide where less than 5\% variation occurred ($<$ 1.96 \micron\/,
{1.99 -- 2.03 \micron\/} and 2.04 -- 2.08 \micron\/).  For (90482)
Orcus, a spectrum of solar analogue HD 102196 collected at an airmass
of 1.24 was used for calibration purposes.  Photometric standards for
(90377) Sedna were not taken on UT date December 27 due to instrument
failure shortly after collection of the science data, with severe
weather during subsequent nights.  A rough photometric calibration was
made using images of UKIRT Faint Standard FS 7 collected on UT dates
December 6 and 7, the temporally closest available photometric
standard, resulting in a $K$ zeropoint of 23.36, consistent with
previous telescope performance.  Mirror resurfacing occurred before
observations of (90482) Orcus, resulting in a fainter $K$ zeropoint of
23.47 for (90482) Orcus.  We find that the magnitudes of (90377) Sedna
and (90482) Orcus were $K \approx 18.9$ and $18.0$, respectively.
Images of (90377) Sedna and (90482) Orcus were compared to field stars
for the detection of possible companions.  No evidence of elongation
was apparent in the 8 minutes of 0\farcs 31 $K$ acquisition images for
(90377) Sedna and 3 minutes of 0\farcs 46 acquisition images for
(90482) Orcus.

\section{Data Reduction}

\begin{figure}
\plotone{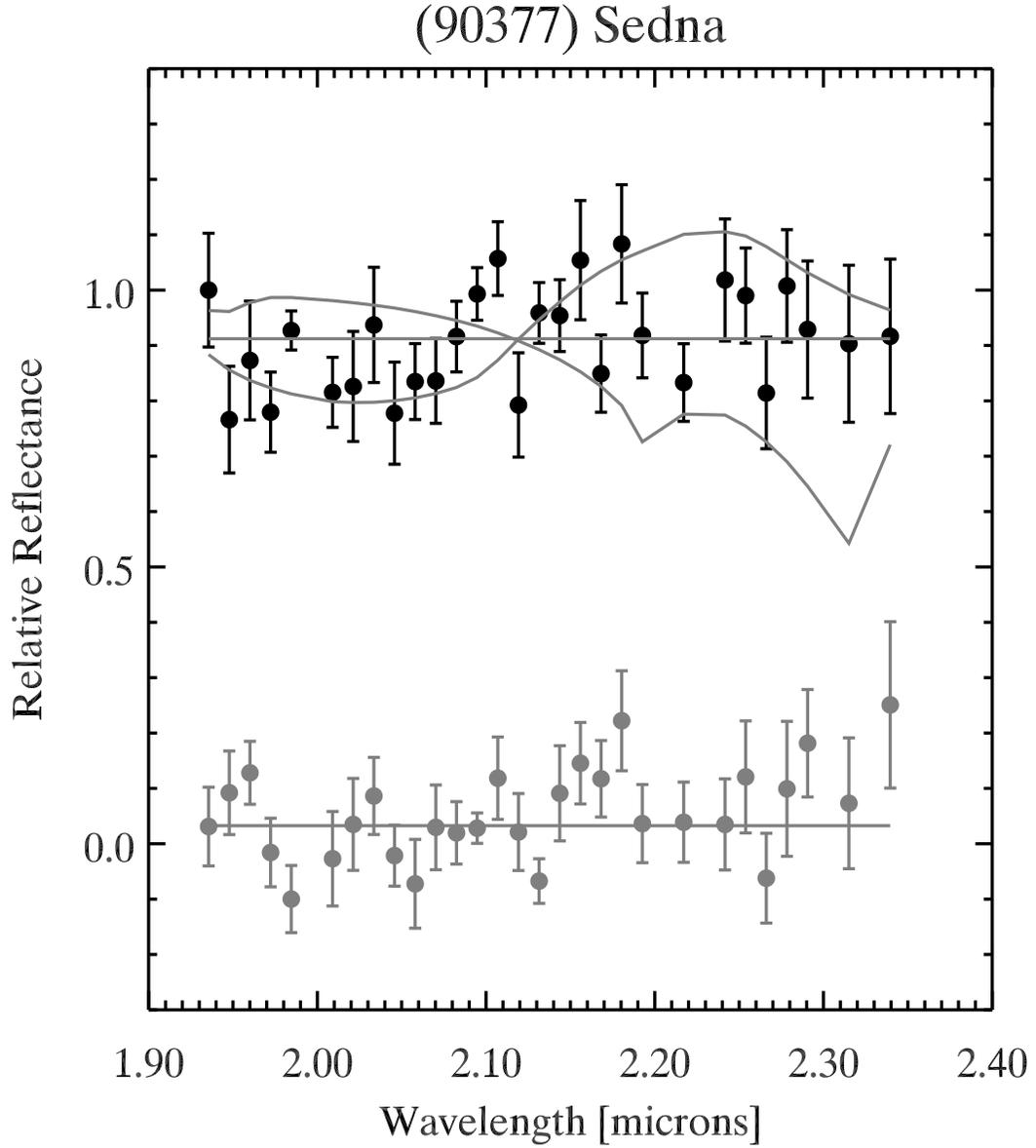} \figcaption{Above is the relative reflectance
  spectrum of (90377) Sedna (black circles) and the spectrum of the
  nearby sky (gray circles).  Gray curves are the model $3 \sigma$
  upper limits to the surface fraction of water ice (smooth gray line)
  and methane ice (jagged gray line).  Surface fractions that cause
  more absorption than the indicated lines are ruled out by our
  observations at the $3 \sigma$ level.  The model shown is for 100
  \micron\/ diameter particles.  Spectrum error bars are computed from the
  reproducibility of the spectral data in each spectral point.
\label{sednaspec}}
\end{figure}

\begin{figure}
\plotone{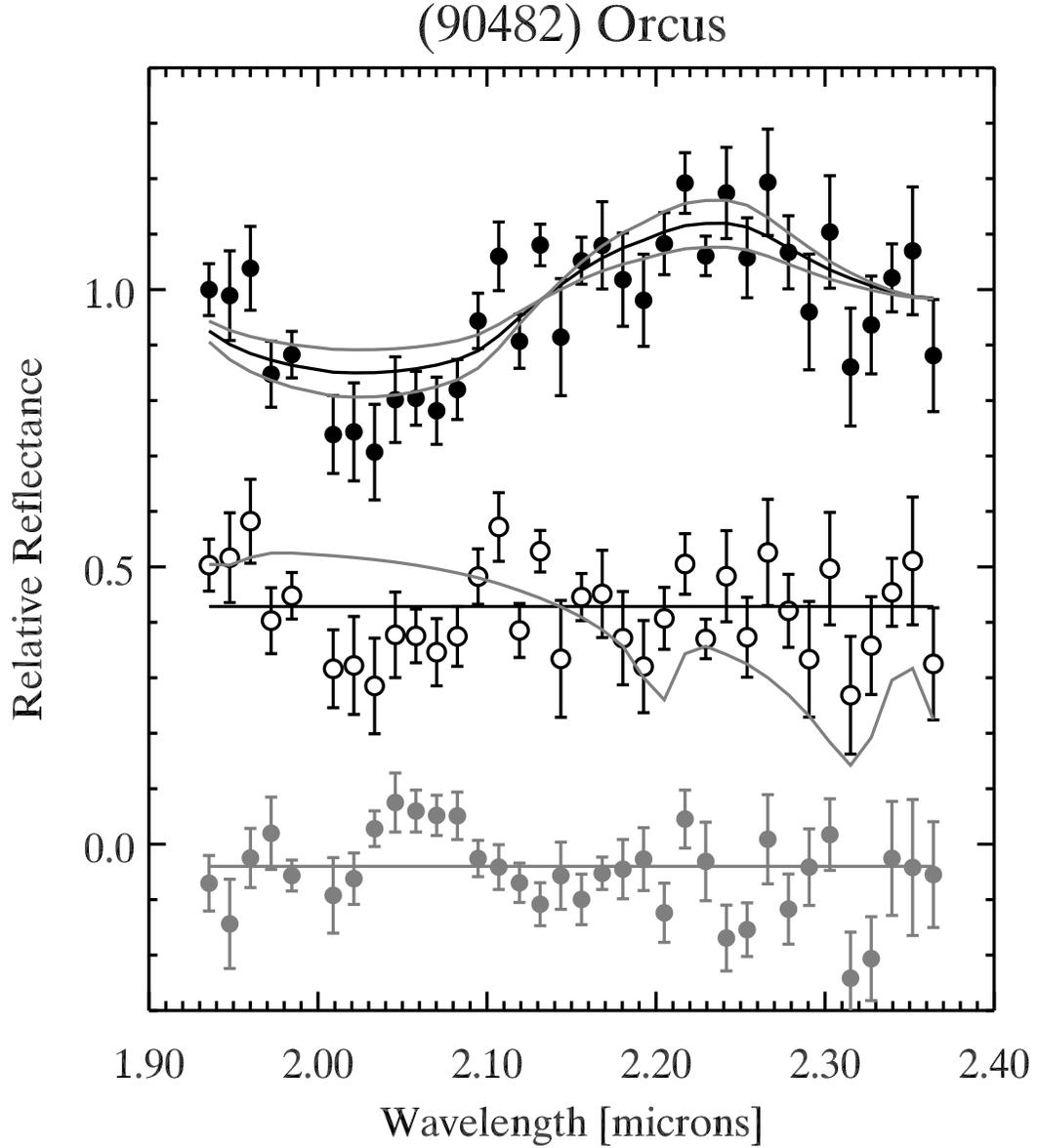} \figcaption{Above is the relative reflectance
  spectrum of (90482) Orcus (black filled circles) and the spectrum of
  the nearby sky (gray circles).  The top spectrum (black filled
  circles) shows the best-fit water ice model and $1\sigma$ limits on
  the best-fit model.  The middle spectrum shows the residual model of
  (90482) Orcus after subtracting the best-fit water ice model (hollow
  circles, offset vertically for clarity).  The gray line on the
  middle spectrum illustrates the 3 $\sigma$ methane ice model.  Any
  greater amount of methane is ruled out by our observations.  The
  model shown is for 100 \micron\/ diameter particles.  Spectrum error
  bars are computed from the reproducibility of the spectral data in
  each spectral point.\label{dwspec}}
\end{figure}

Basic data reduction procedures were followed using the Interactive
Data Language (IDL), produced by Research Systems, Inc.  Processing
included standard procedures for infrared observations such as bad
pixel replacement, pairwise sky subtraction, flat fielding,
combination of science exposures into a single spectrum, wavelength
calibration, and spectrum rectification.  Final processing steps
included spectrum extraction, residual sky subtraction, division by
solar analogue spectra to produce a reflectance spectrum, and
averaging in the spectral direction to increase the signal to noise
ratio at the expense of spectral resolution.  Since NIRI produces
spectra at much higher detector resolution ($7.04 \times 10^{-4}$
\micron\//pixel) than typical solid state molecular transitions
expected for outer solar system objects (0.025 \micron\/), spectral
data were binned to a resolution of 0.01 \micron\/ (14 pixels) by
simple boxcar average.  Errors in spectral flux were estimated for
each wavelength range by measuring the standard error of the pixels
incorporated into the wavelength range.  This estimate is valid for
broad features such as those observed, but will tend to overestimate
errors in narrow line regions, which were not apparent in our spectra.
Overall, the spectrum of (90377) Sedna appears featureless.  The
spectrum of (90482) Orcus in contrast shows strong water ice
absorption near 2.0 \micron\/ and 2.35 \micron\/.  The acquired
reflectance spectra of (90377) Sedna and (90482) Orcus appear in
Figures~\ref{sednaspec} and \ref{dwspec} with 100 \micron\/ grain
diameter models overplotted.

\section{Hapke Model}

Surface properties of the objects were explored using a bidirectional
reflectance model \citep{1993tres.book.....H}.  The reflectance
spectra of (90377) Sedna and (90482) Orcus were compared to
reflectance spectra generated using the Hapke model for methane and
water ice.  For (90377) Sedna, upper limits were placed on the surface
fraction of volatile ices that could be present.  For (90482) Orcus,
which shows strong water ice absorption, the surface fraction of water
ice was estimated, and an upper limit to the fraction of the surface
covered in methane ice was estimated.  Because the albedos of (90377)
Sedna and (90482) Orcus are not known, only relative reflectance
spectra can be determined.  Thus, abundances of compounds which are
relatively featureless in the K-band cannot be quantified, nor can
compounds with very narrow features in the K-band which would not be
seen in our low resolution spectra.  Compounds likely to exist on
outer solar system surfaces which fall into these categories include
carbon black and Titan tholins.  The albedo of the body is considered
a free parameter in our models as the addition of substances such as
carbon black can easily change the albedo without changing the
normalized reflectance observed in this experiment.

Our model produces a geometric albedo (also known as physical albedo
$A_{p}$) for each body, which is the ratio of a brightness of a body
at a phase angle of $g = 0$ to the brightness of a Lambert disk of the
same size and distance of the body observed at opposition.  We model
the physical albedo at zero phase angle using the following equation
from \cite{1993tres.book.....H}
\begin{equation}
A_{p} \simeq r_{0} ( \frac{1}{2} + \frac{1}{6} r_{0}) + \frac{w}{8} [ ( 1
    + B(0)) p(0) - 1],
\end{equation}
where $r_{0}$ is the diffusive reflectance, $p(0)$ is the volume
angular-scattering function, $w$ is the volume single-scattering
albedo, and $B(0)$ is the amplitude of the opposition effect.  The
quantity $r_{0}$ is related to $w$ through the introduction of $\gamma
= \sqrt{1-w}$ such that $r_{0} = (1-\gamma)/(1+\gamma)$.  Under the
assumption of large particles with only one substance present on the
microscopic level, $w = Q_{s}$, where $Q_{s}$ is the scattering
efficiency excluding diffraction.  We have considered only macroscopic
mixtures of components as the differences between models computed for
microscopic and macroscopic mixing are too small to be significant in
our low resolution and low signal-to-noise spectra.  We use the
internal-scattering model to compute $Q_{s}$
\citep{1993tres.book.....H} from the real refractive index $n$, the
imaginary refractive index $k$ and the absorption parameter $\alpha$,
described below.  We did not apply any adjustment to correct from the
phase angle of observation ($g = 0 \fdg 47$ and $0 \fdg 93$ for
(90377) Sedna and (90482) Orcus respectively) to the zero phase angle
modeled for the physical albedo.  Correcting for phase angle would
have made a very minor change to our computed surface fractions, $<
5\%$ of the abundances computed for very deep absorptions such as
those found near pure methane ice transitions.  The opposition effect
of known KBOs was considered in the context of the Hapke model as it
is an input parameter even when modeling zero phase angle as we did.

\subsection{The Opposition Effect}

To date, very little is known about the phase function of the outer
solar system bodies such as the KBOs.  The most comprehensive study to
date was that of \cite{2002AJ....124.1757S}, who found that the KBOs
exhibited remarkable uniformity of opposition surges within 2\arcdeg
of opposition.  All of the seven KBOs studied became darker by $\sim
0.15$ mag per degree from opposition, a trend that was linear with
phase angle up to 2 degrees, the maximum phase of a distant KBO.  This
is a relatively large opposition effect compared to other solar system
objects such as the asteroids \citep{1989aste.conf..524B}.  Any
believable reflectance model of KBO surfaces must reproduce this trend
and in particular must reproduce the very strong (90482) Orcus phase
function of $\sim 0.20$ mag per degree \citep{2004DPS....36.0302R},
which we have adopted in this work.  We considered several different
particle angular-scattering functions $p(g)$ of increasing complexity
as defined in \cite{1993tres.book.....H}: a simple isotropic
scattering function ($p(g) = 1$), a Lambert function, a
Lommel-Seeliger function and a double Henyey-Greenstein function with
two free parameters.  All of the functions were able to reproduce the
phase function found by \cite{2002AJ....124.1757S} with an appropriate
choice of $B(0)$ and parameters of the various models.  For all
models, a very low density surface layer was required, with a filling
factor of only $\phi \sim 5\%$, corresponding to an opposition surge
half-width of about 1 degree.  Such a low filling factor is consistent
with ``fluffy'' material comprising the top visible layer of KBO
surfaces.  For the double Henyey-Greenstein function, values of
$b=0.2$ and $c=1.0$ fit the phase function best, and are typical of
materials with a {\it high} density of internal scatterers within the
particles, characteristic of natural ices with many inclusions and
cracks.  Henyey-Greenstein parameters consistent with a {\it low}
density of internal scatterers fit the observed phase function poorly
and were rejected.  We chose the simplest Hapke grain model parameters
that produced adequately steep opposition effects, namely isotropic
scatterers ($p(g) = 1$) with $B(0) = 2.5$.  This produced opposition
effects of a similar magnitude to those found by
\cite{2002AJ....124.1757S} under a wide variety of assumed albedos.
Our $B(0)$ assumption corresponds to a 20\% increase in flux near
opposition over the typical $B(0) = 1$ assumption used in other Hapke
models.  Grain size uncertainty exceeds opposition effect
uncertainties in all model calculations.

\subsection{Grain Sizes}

\begin{figure}
\plotone{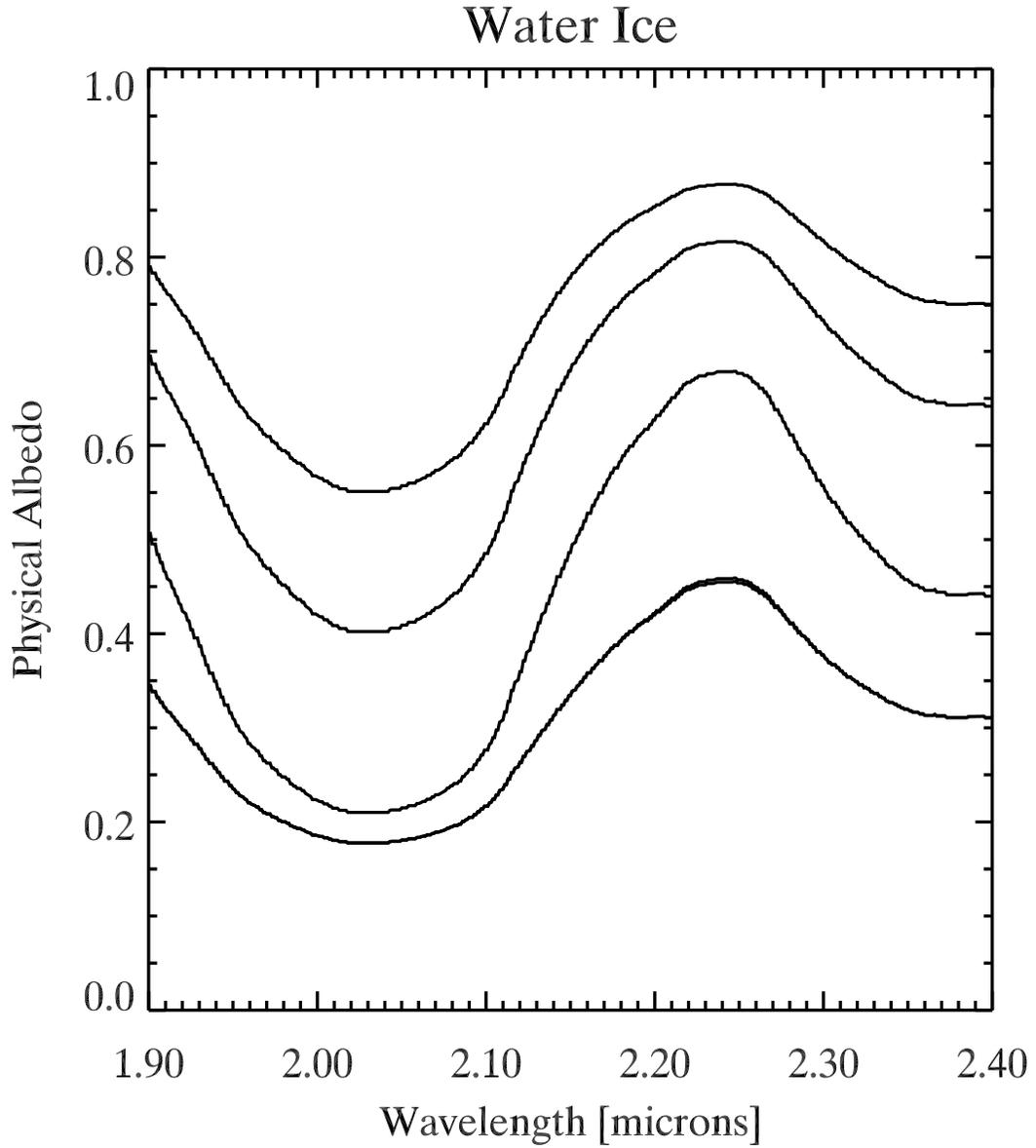} \figcaption{Pictured is the physical
  albedo of a body composed of pure water ice as computed by our Hapke
  model for (top to bottom) grain sizes 10 \micron\/, 25 \micron\/,
  100 \micron\/, 1 cm and 10 cm.  Note that the 1 cm and 10 cm models
  are nearly indistinguishable.  See text for further details of the
  model.
\label{watericemodels}}
\end{figure}

\begin{figure}
\plotone{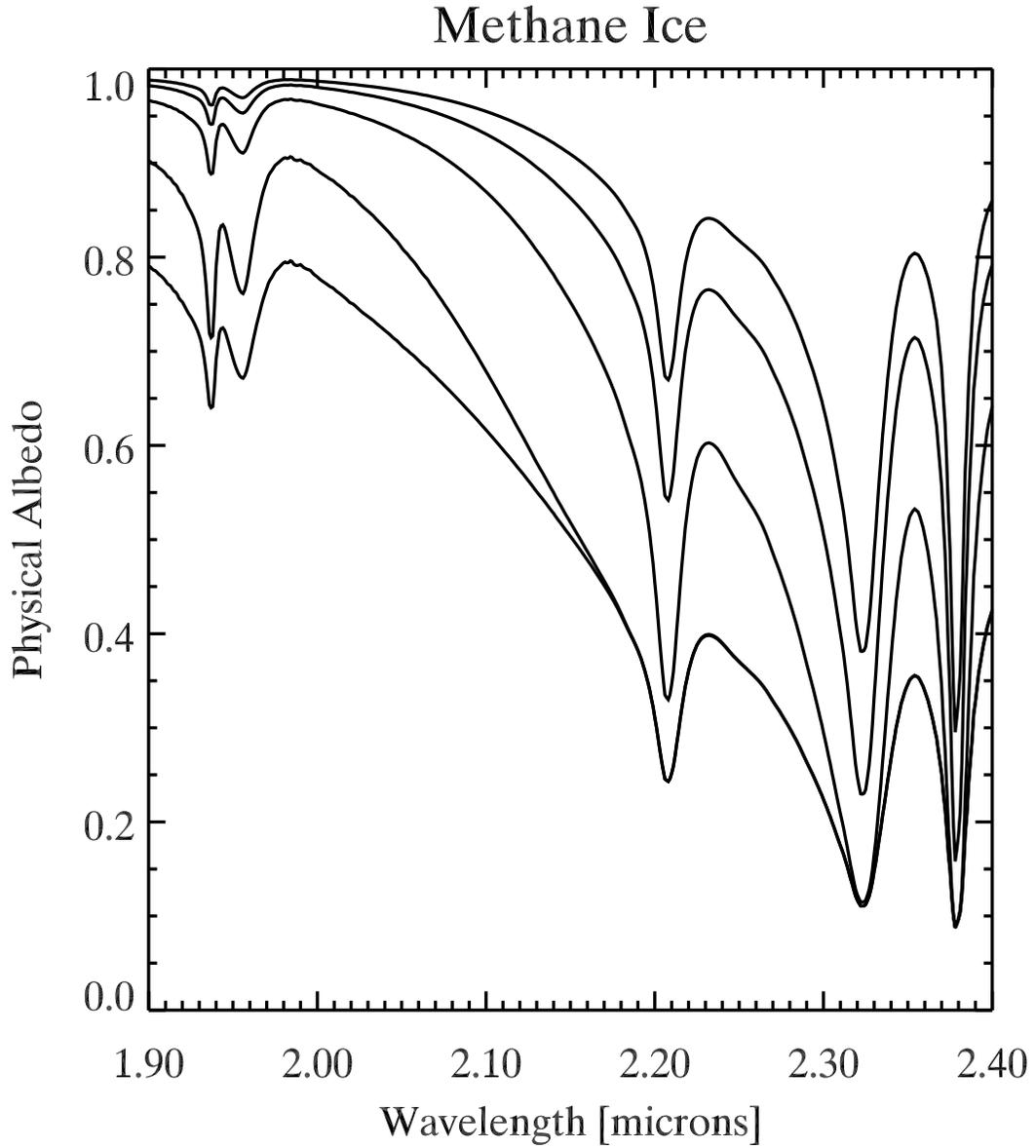} \figcaption{Pictured is the physical
  albedo of a body composed of pure methane ice as computed by our
  Hapke model for (top to bottom) grain sizes 10 \micron\/, 25
  \micron\/, 100 \micron\/, 1 cm and 10 cm.  See text for further
  details of the model.\label{methaneicemodels}}
\end{figure}

We modeled a range of grain sizes since the grain sizes of particles
on (90377) Sedna and (90482) Orcus are not known.  Recent models of
Pluto and Charon used grain diameters of 75 \micron\/ to 1 cm
\citep{1997plch.book..221C} and recent modeling of Triton used grain
sizes of 100 \micron\/ to 9 cm \citep{1999Icar..139..159Q}.  We
modeled spherical particles with diameters of 10 \micron\/, 25
\micron\/, 100 \micron\/, 1 mm and 1 cm, which span a wide range of
absorption in the modeled reflectance spectra.  These models are shown
in Figures~\ref{watericemodels} and \ref{methaneicemodels}.  The
different grain size models change the overall albedo, but they do not
significantly change the shape of the absorptions for water ice and
they only slightly change the shape of the methane bands at the low
resolution observed.  The grain sizes affect the amount of ice
required to produce the observed absorption as larger grains in
general have much more absorption per unit area than smaller grains.
Our relative reflectance model is generally most sensitive to grains
of intermediate size where contrast between different wavelengths in
the $K$ band is greatest because absorption is strong but not yet
saturating, particularly for methane.  However, it should be noted
that the contrast due to absorption in the $K$ band is not a simple
linear function of grain size, so the preceding statement is not
strictly true in all cases.

\subsection{Optical Constants}

Optical constants used in the Hapke model were culled from the
available literature in temperature ranges that are typical of outer
solar system bodies ($< 40$ K).  If no such low temperature laboratory
results were available, higher temperature constants were used.  For
the water ice model, real indices of refraction were obtained from
\cite{1984ApOpt..23.1206W} (266 K) and wavelength dependent absorption
coefficients for hexagonal water ice at 20 K temperatures were found
in \cite{1998JGR...10325809G}.  For methane ice, the real refractive
index was assumed to be uniformly 1.32 over the $K$ band, consistent
with the assumptions of \cite{1991JGR....9617477P}.  The wavelength
dependent absorption coefficient for methane at 30 K was found in
\cite{2002Icar..155..486G}.  For both chemicals, the
wavelength-dependent imaginary index of refraction $k$ was estimated
from the absorption coefficient $\alpha(\lambda)$ using the formula
$k(\lambda) = \alpha(\lambda) \lambda / 4 \pi$, where $\lambda$ is the
wavelength of interest, as used in \cite{1999Icar..139..159Q}.

\subsection{The Neutral Absorber \label{mixalbedo}}

In our models, the overall $K$-band albedo of the body is treated as a
free parameter, $p$.  Only two components are used, that of either
water ice or methane ice (depending on which is being fit), and that
of an absorber neutral in the $K$-band with an albedo chosen to
produce the selected $K$-band albedo $p$.  The albedo of the neutral
absorber depends on the surface fraction $f$ of the body covered with
ice and the mean model $K$-band albedo $p_m$, as computed in the
current ice model (Figures~\ref{watericemodels} and
\ref{methaneicemodels}).  In algebraic terms, the neutral absorber has
albedo $(p-p_m)/(1-f)$.  Physical albedo models requiring neutral
absorber albedos larger than unity were rejected.  There was no
specific chemical composition associated with the neutral absorber.
However, laboratory tholins are mostly neutral in the $K$-band and
show a wide range of reflectances from 0.0 to 0.4
\citep{2004Icar..168..158R}.

\subsection{Model Limitations}

The largest limitations to our models are the unknown $K$ band albedos
of (90377) Sedna and (90482) Orcus.  This is due primarily to the
faintness of the targets in the submillimeter regime which makes it
difficult to obtain thermal observations.  Typically, thermal
observations can be combined with visible photometry to yield albedos,
such as has been done for Varuna and (55565) 2002 $\rm AW_{197}$
\citep{2001Natur.411..446J, 2002DPS....34.1703M}.  This, however can
only be done for a few of the largest bodies and as yet, (90377) Sedna
and (90482) Orcus have not been detected at thermal wavelengths (upper
limits have been placed, however, on (90377) Sedna by
\cite{2004ApJ...617..645B}).  Because of the unknown $K$ band albedos,
we have computed our surface reflectance models for a variety of
assumed albedos.  The second largest model limitation is that the
typical diameters of the surface grains are unknown.  Constraining
grain sizes is quite difficult without very high signal-to-noise
spectra, a known albedo and the presence of narrow absorptions.  As
mentioned previously, in all models, grains of different compositions
were assumed to occupy different parts of the object disk --- all
Hapke modeling assumed homogeneous materials.  There is an additional
limitation to the observations presented here, namely that only one
epoch of observations was conducted.  True planet-wide abundances may
depart from the analysis presented here if the surfaces of either
(90482) Orcus or (90377) Sedna have large inhomogeneities on
hemispherical scales.

\section{(90377) Sedna and (90482) Orcus}

\begin{figure}
\plotone{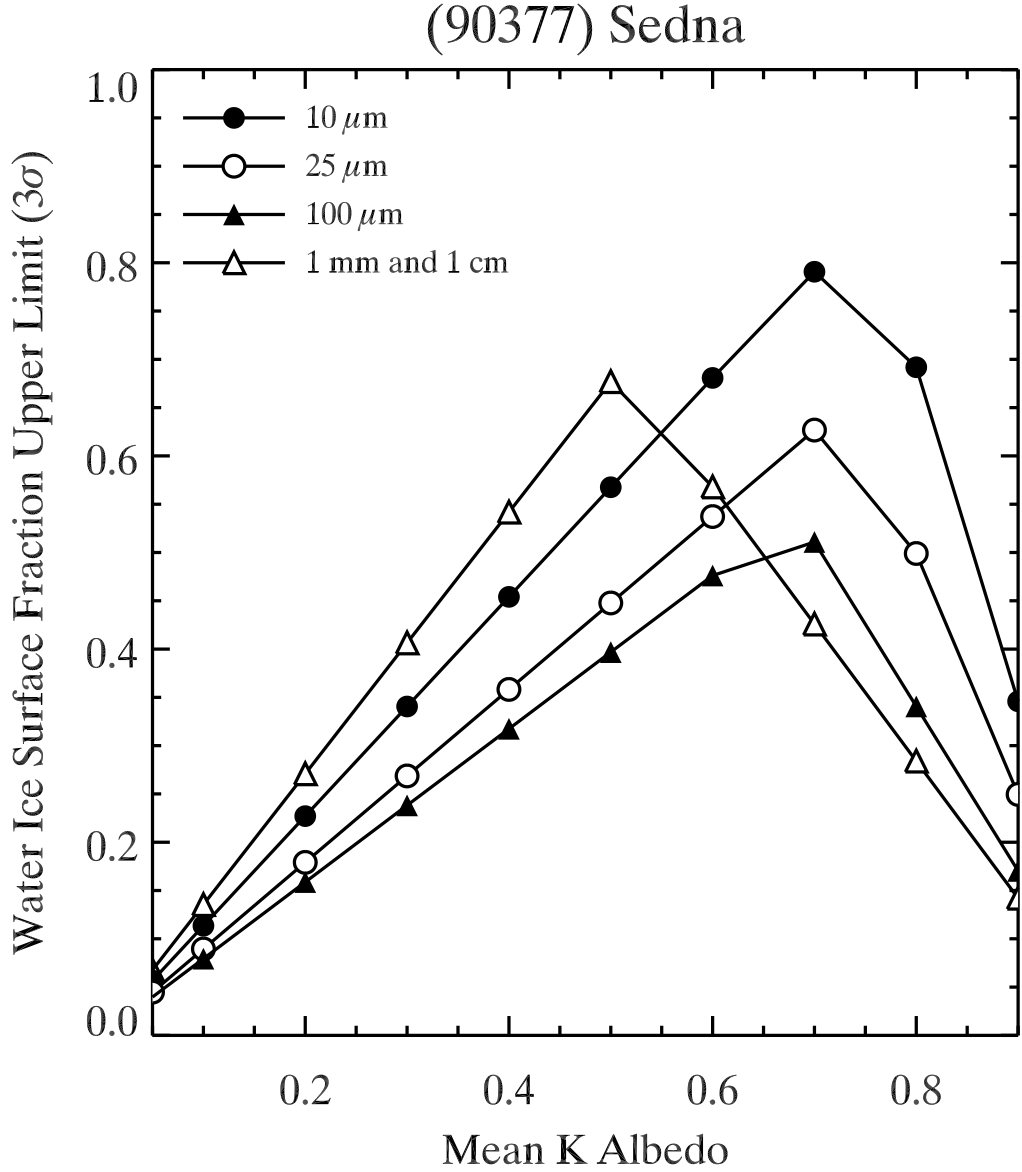} \figcaption{Upper limits ($3 \sigma$) to the
  surface fraction of water ice for (90377) Sedna under a variety of
  assumed mean albedos and grain diameters.  To $3 \sigma$ confidence,
  the surface of (90377) Sedna must be covered by less than 70\% water
  ice for grain diameters 25 \micron\/ or larger.
\label{sednawater}}
\end{figure}

We use the apparently featureless near-infrared reflectance spectrum
of (90377) Sedna to place upper limits on the surface fractions
covered by water ice and methane ice under a wide variety of assumed
albedos.  We characterize the surface of (90377) Sedna with a Hapke
model as described above with three free parameters.  The first two
free parameters are the assumed $K$ band albedo and the grain
diameter.  The third free parameter is the fraction of either water
ice or methane ice.  The best-fit model for each surface composition
was determined by minimizing the $\chi^{2}$ statistic.  For (90377)
Sedna, first water-ice models were considered.  Only the fraction of
water ice was considered as a free parameter for a given grain size
and K-band albedo.  The $\chi^{2}$ statistic was minimized for the
selected grain size and K-band albedo (for all grain sizes and
albedos, the reduced $\chi^{2}$ was between 1.19 and 1.22).  Since no
absorption features were apparent at the noise level of the observed
spectrum, our best-fit $\chi^{2}$ was statistically equivalent to zero
water ice fraction.  We estimated the $3 \sigma$ upper limit to the
amount of water ice that could be present on (90377) Sedna by using an
$F$ test \citep{bevington}.  In the $F$ test, the $\chi^{2}$ computed
for zero water ice was compared to the $\chi^{2}$ computed from
successively larger surface fractions of water ice.  When the
perturbations were so great that the $F$ test signaled a $3 \sigma$
statistical change, this point was marked as the $3 \sigma$ upper
limit for water ice.  This procedure was repeated for all grain sizes
and all assumed K-band albedos with results summarized in
Figure~\ref{sednawater}.  Results appear in their entirety in
Table~\ref{sednatable}.

\begin{figure}
\plotone{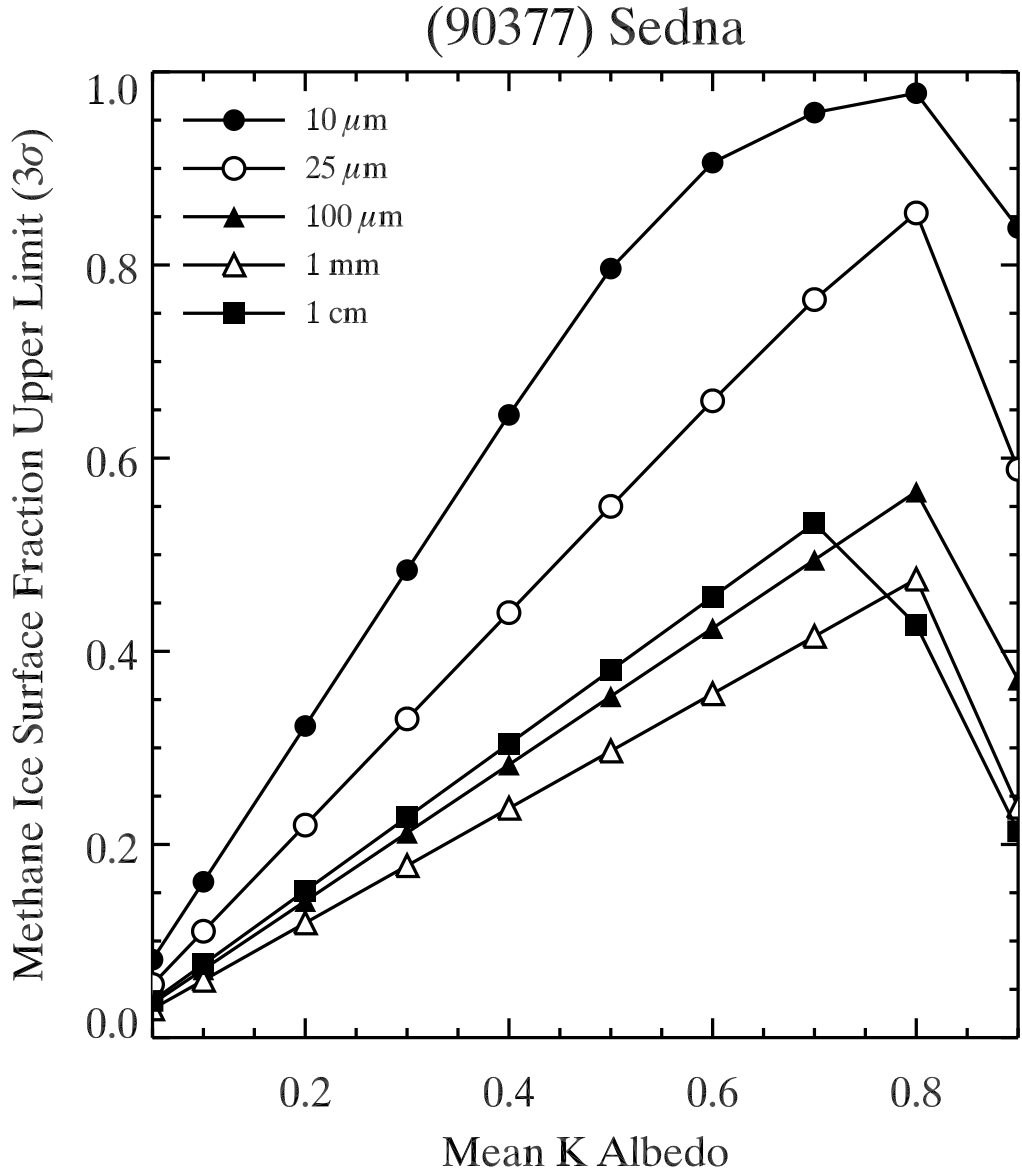} \figcaption{Upper limits ($3 \sigma$) to the
  surface fraction of methane ice for (90377) Sedna under a
  variety of assumed mean albedos and grain diameters.  Assuming grain
  diameters 100 \micron\/ or larger, we find that to $3 \sigma$
  confidence the surface of (90377) Sedna must be covered by less
  than 60\% methane ice. \label{sednamethane}}
\end{figure}

An identical procedure was performed for methane ice on (90377) Sedna
as no methane absorption was apparent either.  Best fit $\chi^{2}$
models were statistically equal to zero methane ice fraction as was
true for the water ice case.  The $\chi^{2}$ for a model with zero
methane ice ($\chi^{2}$ was between 1.19 and 1.22 for all grain sizes
and albedos) was compared to that of increasing fractions of methane
ice until a $3 \sigma$ limit was indicated by an $F$ test.  This
procedure was repeated for all grain sizes and K-band albedos
considered and is summarized in Figure~\ref{sednamethane} and
Table~\ref{sednatable}.  For grain diameters 25 \micron\/ or larger
and all albedo combinations, the surface fraction of water ice on
(90377) Sedna must be less than 70\% or it would be detected in our
observations at the greater than $3 \sigma$ significance level.
Similarly, for methane ice, we find in Figure~\ref{sednamethane} that
for moderate to large grains (diameters 100 \micron\/ or larger), the
surface fraction of methane ice must be less than 60\% for (90377)
Sedna or methane would have been detected in our observations at
greater than $3 \sigma$ significance.

\begin{figure}
\plotone{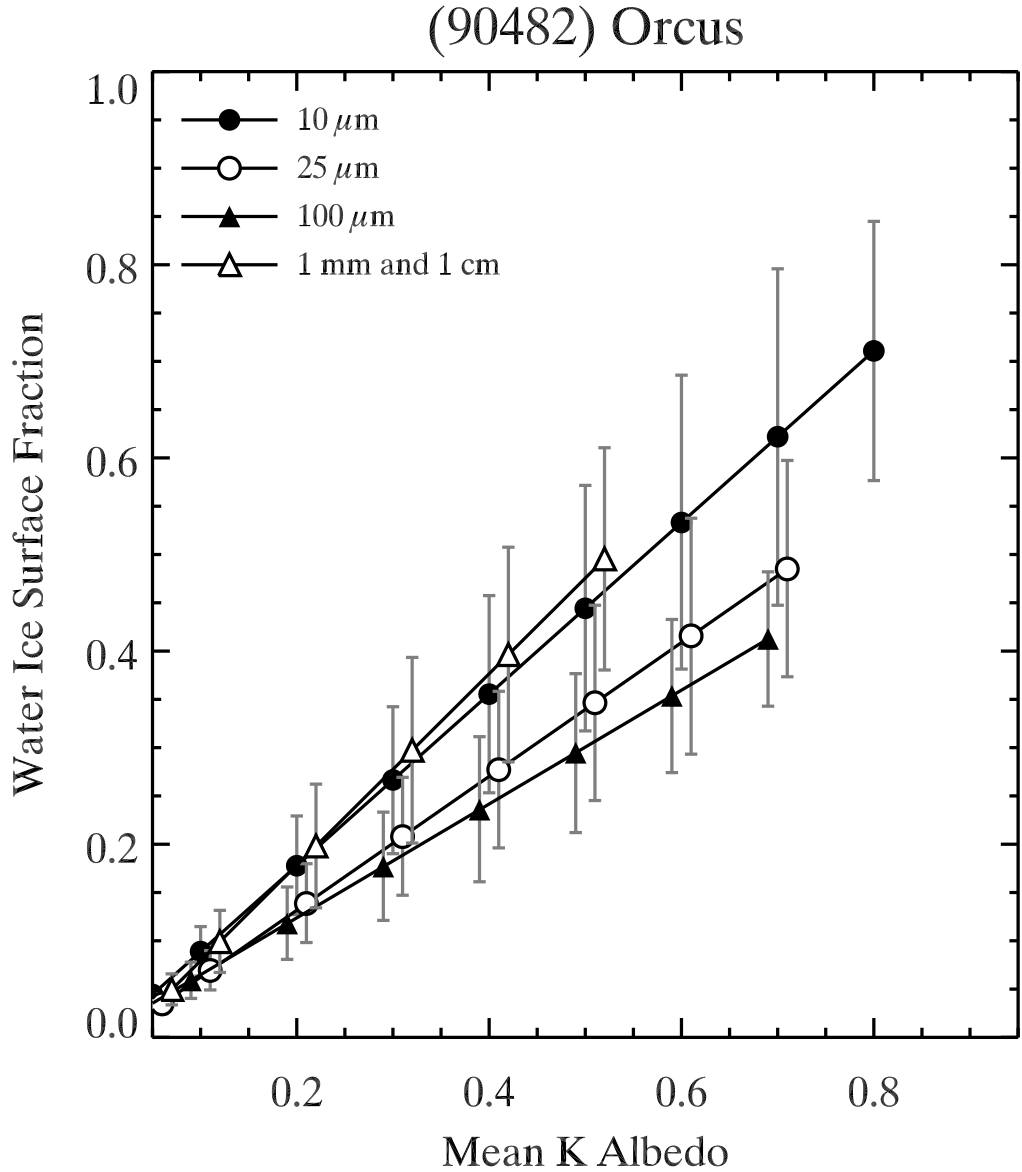} \figcaption{Best fit to the surface fraction of
  water ice for (90482) Orcus under a variety of assumed mean albedos and
  grain diameters (error bars are $1 \sigma$).  Points that do not
  appear are not physically plausible as they require a greater than
  unity albedo for the surface component that is not water ice.
  Unless grains are smaller than 25 \micron\/, the surface of Orcus
  must be covered by less than 50\% water ice.
\label{dwwater}}
\end{figure}

For (90482) Orcus a similar procedure was adopted, with minor
modifications since water-ice was apparent in the spectrum.  Since
water ice was detected, all water ice calculations estimate the $1
\sigma$ error bars on the detection, rather than the $3 \sigma$ upper
limits computed above for (90377) Sedna.  The $\chi^{2}$ statistic was
minimized using the fraction of water ice as a free parameter for the
selected grain size and K-band albedo (for all grain sizes and
albedos, the reduced $\chi^{2}$ was between 1.32 and 1.54).  The
reported $1 \sigma$ limits on the surface fraction of water ice was
again computed by using an $F$ test.  In the $F$ test, the best-fit
$\chi^{2}$ was compared to the $\chi^{2}$ computed from perturbing the
surface fraction of water ice from the best-fit values.  When the
perturbations were so great that the $F$ test signaled a $> 1 \sigma$
statistical change, this point is marked as the $1 \sigma$ error bar
limit.  This procedure was repeated for all grain sizes and all
assumed K-band albedos with results summarized in Figure~\ref{dwwater}
and a displayed in their entirety in Table~\ref{dwtable}.

\begin{figure}
\plotone{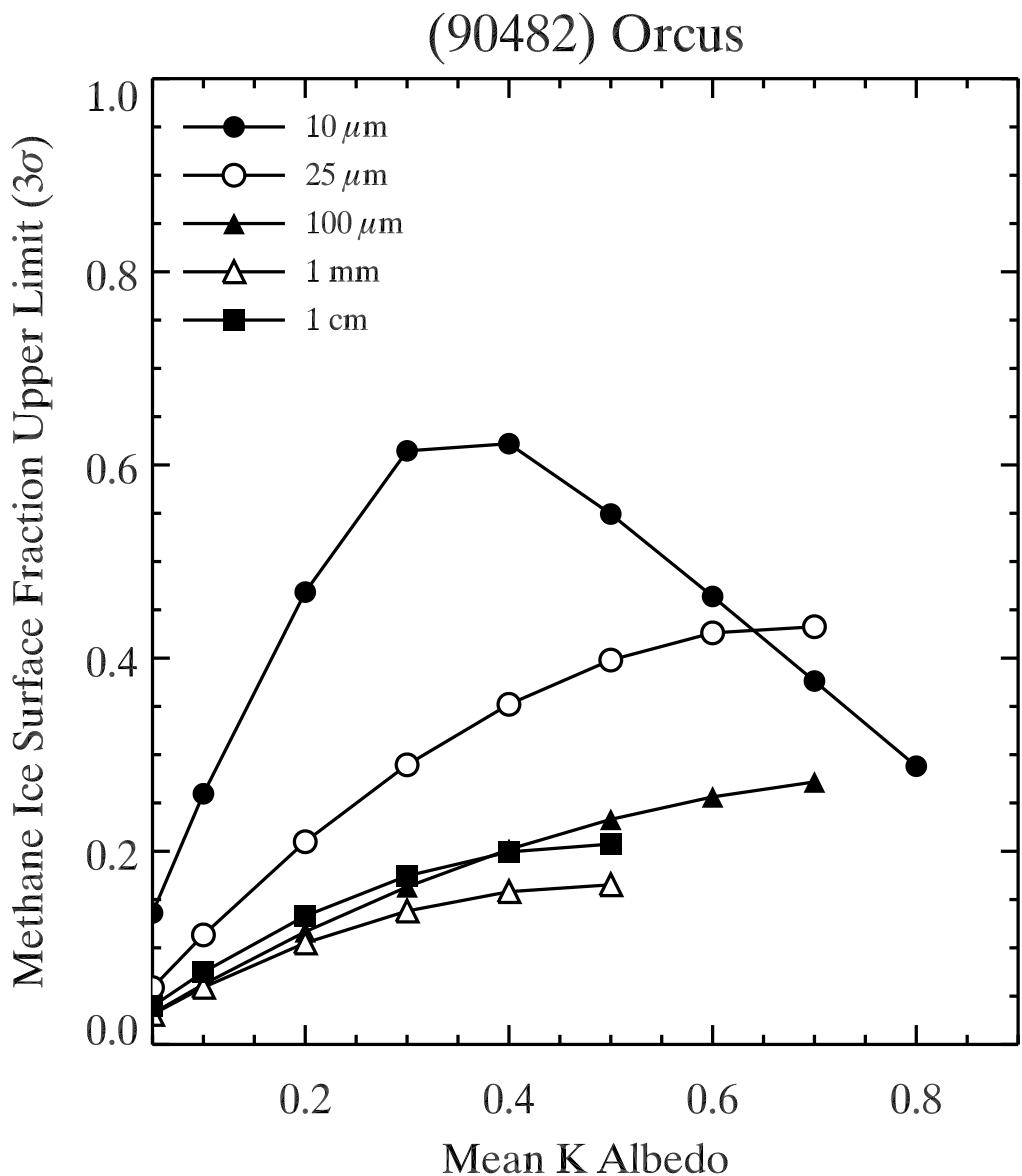} \figcaption{Upper limits ($3 \sigma$) to the
  surface fraction of methane ice for (90482) Orcus under a variety of
  assumed mean albedos and grain diameters after subtraction of the
  best-fit water ice spectrum.  Points that do not appear are not
  physically plausible as they require a greater than unity albedo for
  the non-water ice surface components.  For all models with grains
  $>$ 25 \micron\/, the surface of (90482) Orcus must have less than
  30\% area coverage of methane ice.
\label{dwmethane}}
\end{figure}

The $3 \sigma$ upper limits for methane ice were computed after
subtracting the best-fit water ice model spectrum from (90482) Orcus.
The residual spectrum was then fit with a methane ice model combined
with a neutral absorber.  As for (90377) Sedna, the surface fraction
of methane required to produce the best-fit $\chi^{2}$ did not deviate
significantly from zero, so a neutral absorber was used as our
best-fit spectrum (reduced $\chi^{2}$ was between 1.23 and 1.54
depending on the grain size and K-band albedo assumed).  The $3
\sigma$ upper limit on the amount of methane ice that could be present
was estimated using the $F$ test.  The fraction of methane ice was
increased until the $F$ test computed comparing the new $\chi^{2}$ to
that of zero methane ice $\chi^{2}$ showed a $3 \sigma$ deviation
yielding our $3 \sigma$ upper limit for the amount of methane ice that
could be on the surface of (90482) Orcus.  Again, this procedure was
repeated for all grain sizes and all assumed K-band albedos to produce
Figure~\ref{dwmethane} and Table~\ref{dwtable}.
Figure~\ref{dwmethane} indicates that methane ice must be restricted
to less than 30\% of the surface of (90482) Orcus unless grains are
smaller than 100 \micron\/.  The best-fit models for water ice on the
surface of (90482) Orcus show that less than 50\% of the surface is
covered with water ice if grains are 25 \micron\/ or larger
(Figure~\ref{dwwater}).  Note that there is an upper limit to the $K$
band albedo for (90482) Orcus of about 0.7 for most grain models.
Models with $>$ 0.7 $K$-band albedo would require a mixing albedo
greater than unity in the parts of (90482) Orcus not covered by water
ice and are thus physically implausible.


\section{Comparison to Known Bodies}

Due to its size and very low temperature environment, it might be
reasonable to expect the near infrared observations of the surface of
(90377) Sedna to be similar to that of either Pluto (dominated by
methane ice absorption) or Charon (dominated by water ice absorption).
Figure~\ref{sednawater} rules out a Charon-like surface immediately as
Charon has a $K$ band albedo of $\sim 20\%$ and a surface fraction of
water ice of 80\% assuming grains of diameter 100\micron\/
\citep{1996Icar..119..214R}.  Figure~\ref{sednamethane} is not
directly comparable to recent models of Pluto's methane
\citep{1997plch.book..221C}, as most Pluto models use methane dissolved
in nitrogen, a model not considered in this work.  However, a direct
$F$-test comparing (90377) Sedna's spectrum to that of Pluto rules out
the possibility that the surface of (90377) Sedna could be identical
to that of Pluto at the $3\sigma$ level.

The dissimilarity of both the Pluto and Charon spectrum with (90377)
Sedna is somewhat unexpected, as these bodies are in principle similar
to (90377) Sedna in terms of size and temperature ($\sim 40$K for
Pluto/Charon and $\sim 37$K for (90377) Sedna), although not in color
or origin.  We note that the vast majority of Pluto's atmosphere is
expected to be nitrogen, with only trace amounts of methane
responsible for most of the near-infrared color
\citep{1997plch.book..221C}.  Thus, although the spectrum of (90377)
Sedna may be dissimilar to Pluto's, this difference could be primarily
due to the methane abundance, since the nitrogen fraction on (90377)
Sedna is unknown.

Pholus is a prime candidate for a possible compositional analogue for
(90377) Sedna.  Pholus' surface (one of the reddest in the solar
system) and Sedna's are similar color in the visible
\citep{2004DPS....36.0302R}.  The Pholus model is consistent with the
observed (90377) Sedna surface at the signal to noise collected, and
the water ice fraction observed on Pholus (15\% for a 10 \micron\/
model) is well within the limits for (90377) Sedna
\citep{1998Icar..135..389C}.  Recently, \cite{2004DPS....36.0301B}
reported the detection of tholins on (90377) Sedna, consistent with a
Pholus-like composition.  Such tholins could not have been detected in
our $K$-band spectra alone.

The water ice observed on (90482) Orcus exists in moderate amounts.
Assuming an albedo of roughly 20\% to 40\% in the K-band, such as
found on Charon (20\%) and Pluto (30\%), we estimate that the surface
of (90482) Orcus likely has between 15\% and 30\% water ice coverage,
depending on the size of the grains involved.  This is significantly
less coverage than is found on Charon (80\%)
\citep{1996Icar..119..214R}, but similar to the amount found on Triton
\citep[40\%]{2000Icar..147..309C}.  Since we find that the surface of
(90482) Orcus cannot be entirely composed of water ice, it is possible
that future work on (90482) Orcus may uncover more chemical
components.

Although we have quantified the surface fraction of ices for a model K
albedos between 0.05 and 1.0 for completeness, a smaller possible
range of albedos is more probable given our current knowledge of outer
solar system bodies.  The icy Galilean satellites, for instance, have
among the largest amounts of water ice observed in the solar system
yet they have K-band albedos between 0.25 and 0.35 outside water
absorption bands as measured by Cassini VIMS under a variety of phase
angles \citep{2003Icar..164..461B}.  One effect that may artificially
increase the physical albedo of a KBO over that of the Galilean
satellites is that KBOs are always observed near opposition due to
their extreme distances.  However, even on icy Europa, the opposition
surge is found to increase the flux by about 0.40 in the K-band
\citep{2004Icar..172..149S}.  Combining the expected opposition surge
for an icy body (0.40) and reasonable albedos for well-studied icy
bodies (0.25 - 0.35), we expect the upper limit for the K-band albedos
to be about 0.50.  Future works will undoubtedly test this further for
both Sedna and Orcus.  In any case, models have been run for high
K-band albedos, but we believe that the results of the 0.5 or less
models are the most physically relevant.

Using this stricter 0.5 albedo criterion for the possible range of
albedos revises the results listed in the previous section only
slightly.  For (90377) Sedna, less than 40\% of the surface can be
covered with methane ice to $3 \sigma$ while the water ice results are
unchanged.  For (90482) Orcus, the water ice model with maximum
surface fraction shows 50\% $\pm$ 16\% water ice for all grain sizes,
and there is no change to the methane ice results.

\section{Summary and Future Work}

We have obtained a near-infrared spectrum of (90377) Sedna using the
Gemini 8-m telescope.  The spectrum is featureless and consistent with
other solar system objects with low to moderate absorptions in the $K$
band.  Compositions such as those of Pluto and Charon, which are
dominated by very strong methane ice and water ice absorption
respectively can be ruled out with greater than $3 \sigma$ confidence.
No evidence of a companion was found in the 0\farcs 31 $K$ band images
for (90377) Sedna.  Combining these data with observations in the
visible \citep{2004DPS....36.0302R} suggests that (90377) Sedna has a surface of
similar color to Pholus in the visible (presumably due to a Titan
tholin-like compounds), and has moderate, if any, absorption bands in
the $K$ region.

The surface of (90482) Orcus is considerably easier to characterize
than (90377) Sedna due to its brightness and the presence of moderate
water ice features at 2 \micron\/ and 2.4\micron\/.  No binary was
seen in the 0\farcs 46 $K$ band images for (90482) Orcus.  Using the
Hapke models we have produced, we can place crude constraints on the
surface of the bodies studied.  Specifically, we find the following:

(1) To $3 \sigma$ confidence, the surface of (90377) Sedna must be
    covered by less than 70\% water ice under most grain models and
    albedo combinations studied.

(2) Assuming moderate to large grain models (diameters 100 \micron\/
    or larger), the surface of (90377) Sedna must be covered by less
    than 60\% methane ice to $3 \sigma$ confidence.

(3) Unless grain diameters on (90482) Orcus are smaller than 25 \micron\/,
    the surface of (90482) Orcus cannot be covered by more than 30\% methane
    ice.

(4) The best-fit models for grain diameters 25 \micron and larger
    suggest that the water ice surface fraction on (90482) Orcus is
    less than 50\%.  The maximum best-fit surface fraction of water
    ice is 50\% $\pm$ 16\%.

(5) When the albedos of (90377) Sedna and (90482) Orcus are measured,
    the above results will be significanly more constrained as only
    one albedo model need be considered.

(6) Very low-density grain models with filling factors of $\sim 5\%$
    are required to reproduce the strong opposition effect observed
    for KBOs with our Hapke model.

For both bodies, considerably more observations and analysis are
needed.  Constraining the albedos of the bodies would place strong
limits on the surface fraction covered by water ice and methane ice
even given the fact that the grain sizes on the surfaces are unknown.
Additionally, in this work, each object was observed at a single
epoch.  Once the rotation parameters of the bodies are known, it would
be prudent to observe each of these bodies through a complete rotation
as well as to consider additional chemical components as they become
evident in higher signal to noise spectra.

\acknowledgements

We thank C. M. Mountain for granting Director's
Discretionary time for this project.  Alan Hatakeyama's support at the
telescope was greatly appreciated.  Operations at the telescope were
greatly aided with the help of Simon Chan.  Joe Jensen provided
helpful input into constructing the observation sequence.


\begin{deluxetable}{rrrr}
\tablecolumns{4}
\tablecaption{(90377) Sedna}
\tablehead{
\colhead{Grain diameter} & \colhead{Mean $K$} & \colhead {Water Fraction} & \colhead {Methane Fraction} \\
\colhead{[\micron\/]}       & \colhead{Albedo}   & \colhead {Upper Limit  ($3\sigma$)} & \colhead{Upper Limit ($3\sigma$)}}
\startdata
10    & 0.05 &  6\%                  &  8\%                  \\
10    & 0.10 & 11\%                  & 16\%                  \\
10    & 0.20 & 23\%                  & 32\%                  \\
10    & 0.30 & 34\%                  & 48\%                  \\
10    & 0.40 & 45\%                  & 64\%                  \\
10    & 0.50 & 57\%                  & 80\%                  \\
10    & 0.60 & 68\%                  & 91\%                  \\
10    & 0.70 & 79\%                  & 96\%                  \\
10    & 0.80 & 69\%\tablenotemark{a} & 98\%                  \\
10    & 0.90 & 35\%\tablenotemark{a} & 84\%\tablenotemark{a} \\
25    & 0.05 &  4\%                  &  6\%                  \\
25    & 0.10 &  9\%                  & 11\%                  \\
25    & 0.20 & 18\%                  & 22\%                  \\
25    & 0.30 & 27\%                  & 33\%                  \\
25    & 0.40 & 36\%                  & 44\%                  \\
25    & 0.50 & 45\%                  & 55\%                  \\
25    & 0.60 & 54\%                  & 66\%                  \\
25    & 0.70 & 63\%                  & 76\%                  \\
25    & 0.80 & 50\%\tablenotemark{a} & 85\%                  \\
25    & 0.90 & 25\%\tablenotemark{a} & 59\%\tablenotemark{a} \\
100   & 0.05 &  4\%                  &  4\%                  \\
100   & 0.10 &  8\%                  &  7\%                  \\
100   & 0.20 & 16\%                  & 14\%                  \\
100   & 0.30 & 24\%                  & 21\%                  \\
100   & 0.40 & 32\%                  & 28\%                  \\
100   & 0.50 & 40\%                  & 35\%                  \\
100   & 0.60 & 48\%                  & 42\%                  \\
100   & 0.70 & 51\%\tablenotemark{a} & 49\%                  \\
100   & 0.80 & 34\%\tablenotemark{a} & 57\%                  \\
100   & 0.90 & 17\%\tablenotemark{a} & 37\%\tablenotemark{a} \\
1000  & 0.05 &  7\%                  &  3\%                  \\
1000  & 0.10 & 14\%                  &  6\%                  \\
1000  & 0.20 & 27\%                  & 12\%                  \\
1000  & 0.30 & 41\%                  & 18\%                  \\
1000  & 0.40 & 54\%                  & 24\%                  \\
1000  & 0.50 & 68\%                  & 30\%                  \\
1000  & 0.60 & 57\%\tablenotemark{a} & 36\%                  \\
1000  & 0.70 & 43\%\tablenotemark{a} & 42\%                  \\
1000  & 0.80 & 28\%\tablenotemark{a} & 47\%                  \\
1000  & 0.90 & 14\%\tablenotemark{a} & 24\%\tablenotemark{a} \\
10000 & 0.05 &  7\%                  &  4\%                  \\
10000 & 0.10 & 14\%                  &  8\%                  \\
10000 & 0.20 & 27\%                  & 15\%                  \\
10000 & 0.30 & 41\%                  & 23\%                  \\
10000 & 0.40 & 55\%                  & 30\%                  \\
10000 & 0.50 & 68\%                  & 38\%                  \\
10000 & 0.60 & 57\%\tablenotemark{a} & 46\%                  \\
10000 & 0.70 & 43\%\tablenotemark{a} & 53\%                  \\
10000 & 0.80 & 28\%\tablenotemark{a} & 43\%\tablenotemark{a} \\
10000 & 0.90 & 14\%\tablenotemark{a} & 21\%\tablenotemark{a} \\
\enddata
\tablenotetext{a}{Upper limit is a physical limit, any higher fraction
would require a mixing albedo greater than unity.}
\tablecomments{Upper limits to surface fraction of water ice and methane ice
  on (90377) Sedna for a variety of grain sizes and mean $K$
  albedos. }
\label{sednatable}
\end{deluxetable}

\begin{deluxetable}{rrrr}
\tablecolumns{4}
\tablecaption{(90482) Orcus}
\tablehead{
\colhead{Grain diameter} & \colhead{Mean $K$} & \colhead {Water Fraction} & \colhead {Methane Fraction} \\
\colhead{[\micron\/]}       & \colhead{Albedo}   & \colhead {($1\sigma$ errors)} & \colhead{Upper Limit ($3\sigma$)}}
\startdata
10    & 0.05 & $ 4^{+ 1}_{- 1}$\%                  & 14\%                  \\
10    & 0.10 & $ 9^{+ 3}_{- 3}$\%                  & 26\%                  \\
10    & 0.20 & $18^{+ 5}_{- 5}$\%                  & 47\%                  \\
10    & 0.30 & $27^{+ 8}_{- 8}$\%                  & 61\%                  \\
10    & 0.40 & $36^{+10}_{-10}$\%                  & 62\%                  \\
10    & 0.50 & $44^{+13}_{-13}$\%                  & 55\%                  \\
10    & 0.60 & $53^{+15}_{-15}$\%                  & 46\%                  \\
10    & 0.70 & $62^{+17}_{-17}$\%                  & 38\%                  \\
10    & 0.80 & $71^{+ 2}_{-18}$\%\tablenotemark{a} & 29\%                  \\
10    & 0.90 & \tablenotemark{b} & \tablenotemark{b} \\
25    & 0.05 & $ 3^{+ 1}_{- 1}$\%                  &  6\%                  \\
25    & 0.10 & $ 7^{+ 2}_{- 2}$\%                  & 11\%                  \\
25    & 0.20 & $14^{+ 4}_{- 4}$\%                  & 21\%                  \\
25    & 0.30 & $21^{+ 6}_{- 6}$\%                  & 29\%                  \\
25    & 0.40 & $28^{+ 8}_{- 8}$\%                  & 35\%                  \\
25    & 0.50 & $35^{+10}_{-10}$\%                  & 40\%                  \\
25    & 0.60 & $42^{+12}_{-12}$\%                  & 43\%                  \\
25    & 0.70 & $49^{+14}_{-14}$\%                  & 43\%                  \\
25    & 0.80 & \tablenotemark{b} & \tablenotemark{b} \\
25    & 0.90 & \tablenotemark{b} & \tablenotemark{b} \\
100   & 0.05 & $ 3^{+ 1}_{- 1}$\%                  &  3\%                  \\
100   & 0.10 & $ 6^{+ 2}_{- 2}$\%                  &  6\%                  \\
100   & 0.20 & $12^{+ 4}_{- 4}$\%                  & 12\%                  \\
100   & 0.30 & $18^{+ 6}_{- 6}$\%                  & 16\%                  \\
100   & 0.40 & $24^{+ 8}_{- 7}$\%                  & 20\%                  \\
100   & 0.50 & $29^{+ 9}_{- 9}$\%                  & 23\%                  \\
100   & 0.60 & $35^{+11}_{-11}$\%                  & 26\%                  \\
100   & 0.70 & $41^{+12}_{-13}$\%\tablenotemark{a} & 27\%                  \\
100   & 0.80 & \tablenotemark{b} & \tablenotemark{b} \\
100   & 0.90 & \tablenotemark{b} & \tablenotemark{b} \\
1000  & 0.05 & $ 5^{+ 2}_{- 2}$\%                  &  3\%                  \\
1000  & 0.10 & $10^{+ 3}_{- 3}$\%                  &  6\%                  \\
1000  & 0.20 & $20^{+ 6}_{- 6}$\%                  & 11\%                  \\
1000  & 0.30 & $30^{+10}_{-10}$\%                  & 14\%                  \\
1000  & 0.40 & $40^{+13}_{-13}$\%                  & 16\%                  \\
1000  & 0.50 & $50^{+16}_{-16}$\%                  & 17\%                  \\
1000  & 0.60 & \tablenotemark{b} & \tablenotemark{b} \\
1000  & 0.70 & \tablenotemark{b} & \tablenotemark{b} \\
1000  & 0.80 & \tablenotemark{b} & \tablenotemark{b} \\
1000  & 0.90 & \tablenotemark{b} & \tablenotemark{b} \\
10000 & 0.05 & $ 5^{+ 2}_{- 2}$\%                  &  4\%                  \\
10000 & 0.10 & $10^{+ 3}_{- 3}$\%                  &  7\%                  \\
10000 & 0.20 & $20^{+ 6}_{- 6}$\%                  & 13\%                  \\
10000 & 0.30 & $30^{+10}_{-10}$\%                  & 17\%                  \\
10000 & 0.40 & $40^{+13}_{-13}$\%                  & 20\%                  \\
10000 & 0.50 & $50^{+16}_{-16}$\%                  & 21\%                  \\
10000 & 0.60 & \tablenotemark{b} & \tablenotemark{b} \\
10000 & 0.70 & \tablenotemark{b} & \tablenotemark{b} \\
10000 & 0.80 & \tablenotemark{b} & \tablenotemark{b} \\
10000 & 0.90 & \tablenotemark{b} & \tablenotemark{b} \\
\enddata
\tablenotetext{a}{Upper limit is a physical limit, any higher fraction
would require a mixing albedo greater than unity.}
\tablenotetext{b}{Best fit for the fraction of water ice requires a
  mixing albedo greater than unity and is omitted.}
\tablecomments{Upper limits to surface fraction of water ice and methane ice
  on (90377) Sedna for a variety of grain sizes and mean $K$
  albedos. }
\label{dwtable}
\end{deluxetable}



\end{document}